\documentclass[10pt,journal,compsoc]{IEEEtran}

\usepackage{svg}

\usepackage{cite}
\usepackage{amsmath,amsfonts}
\usepackage{xcolor}
\usepackage[inline,shortlabels]{enumitem}
\usepackage{xspace}
\usepackage{multirow}
\makeatletter
\newcommand\dlmu[2][4cm]{\hskip1pt\underline{\hb@xt@ #1{\hss#2\hss}}\hskip3pt}
\makeatother
\usepackage{algorithm}
\usepackage{algorithmic}
\usepackage{booktabs}
\usepackage{inputenc}
\usepackage{footnote}

\usepackage{makecell}
\usepackage{tikz}
\usetikzlibrary{patterns}
\usepackage{caption}
\usepackage{graphicx}
\usepackage{subfigure}
\usepackage{pgfplots}
    \usepgfplotslibrary{groupplots}
    \pgfplotsset{compat=1.9}
\usetikzlibrary{pgfplots.external}
\makeatletter
\tikzset{
    every picture/.style={
        execute at begin picture={
            \let\ref\@refstar
        }
    }
}
\makeatother

\usepackage[T1]{fontenc}
\usepackage{url}
\usepackage{hyperref}

\graphicspath{{figures/}}

\newcommand{\eg}{\textit{e.g.},~}
\newcommand{\ie}{\textit{i.e.},~}
\newcommand{\aka}{\textit{a.k.a.}~}
\newcommand{\etc}{\textit{etc}}

\newcommand{\method}{\textit{DiagFusion}\xspace}

\newcommand{\kgshort}{DG\xspace}
\newcommand{\kglong}{dependency graph\xspace}

\newcommand{\figref}[1]{Figure~\ref{#1}}
\newcommand{\secref}[1]{Section~\ref{#1}}
\newcommand{\tabref}[1]{Table~\ref{#1}}

\newcommand{\FTone}{High memory usage}

\newcommand{\FTthree}{Login failure}

\newcommand{\FTfive}{Access denied}

\newlist{rules}{enumerate}{2}
\setlist[rules,1]{
    leftmargin=*, 
    label=\textbf{RULE \arabic*:},
    ref=\textbf{RULE \arabic*}
}
\setlist[rules,2]{label={(\alph*)}}

\newcommand{\Done}{D1\xspace}
\newcommand{\Dtwo}{D2\xspace}

\newcommand{\draft}[1]{#1}
\newcommand{\ddraft}[1]{#1}

\newcommand{\new}[1]{{{#1}}}

\newcommand{\vspacefigup}{\vspace{-6 pt}}
\newcommand{\vspacefigdown}{\vspace{-10 pt}}
\newcommand{\vspacetbup}{\vspace{-6 pt}}
\newcommand{\vspacetbdown}{\vspace{-6 pt}}

\begin{document}
\bstctlcite{BSTcontrol}

\title{Robust Failure Diagnosis of Microservice System through Multimodal Data}

\author{
    Shenglin Zhang, \IEEEmembership{Member, IEEE,} 
Pengxiang Jin,
Zihan Lin,
Yongqian Sun, \IEEEmembership{Member, IEEE,} 
\\
Bicheng Zhang,
Sibo Xia,
Zhengdan Li,
Zhenyu Zhong,
Minghua Ma,
Wa Jin,
Dai Zhang,
Zhenyu Zhu,
\\
Dan Pei, \IEEEmembership{Senior Member, IEEE}

\IEEEcompsocitemizethanks{
     \IEEEcompsocthanksitem
     Y. Sun is the corresponding author.
     P. Jin, Z. Lin, Y. Sun, S. Xia, Z. Li, Z. Zhong, and W. Jin are with Nankai University, Tianjin, China. 
      Email: \{sunyongqian, lzd\}@nankai.edu.cn, \{jinpengxiang, linzihan, xiasibo, zyzhong, 1913173\}@mail.nankai.edu.cn.
    \new{
    \IEEEcompsocthanksitem
     S. Zhang is with the College of Software, Nankai University, Tianjin, China, and also with the Haihe Laboratory of Information Technology Application Innovation (HL-IT), Tianjin, China. 
      Email: zhangsl@nankai.edu.cn}
    \IEEEcompsocthanksitem
    B. Zhang is with Fudan University, Shanghai, China. Email: 22210240069@m.fudan.edu.cn.
     \IEEEcompsocthanksitem
     M. Ma is with Microsoft, Beijing, China. Email: minghuama@microsoft.com.
     \IEEEcompsocthanksitem
     D. Zhang and Z. Zhu are with ZhejiangE-CommerceBank Co., Ltd. Email: \{henry.zd, michael.zzy\}@mybank.cn
     \IEEEcompsocthanksitem
     D. Pei is with Department of Computer Science, Tsinghua University, Beijing, China, and also with Beijing National Research Center for Information Science and Technology. Email: peidan@tsinghua.edu.cn.
     \new{
     \IEEEcompsocthanksitem 
     This work was supported in part by the National Natural Science Foundation of China (Grant No. 62072264, 62272249), and in part by the Natural Science Foundation of Tianjin (Grant No. 21JCQNJC00180).}
}

}

\IEEEtitleabstractindextext{
    \begin{abstract}
        Automatic failure diagnosis is crucial for large microservice systems.
Currently, most failure diagnosis methods rely solely on single-modal data (\ie using either metrics, logs, or traces).
In this study, we conduct an empirical study using real-world failure cases to show that combining these sources of data (multimodal data) leads to a more accurate diagnosis.
However, effectively representing \new{these} data and addressing imbalanced failures remain \new{challenging}.
To tackle these issues, we \new{propose} \method, a robust failure diagnosis approach that uses multimodal data.
It leverages embedding techniques and data augmentation to represent the multimodal data of service instances, combines deployment data and traces to build a \kglong, and uses a graph neural network to localize the root cause instance and determine the failure type.
Our evaluations using real-world datasets show that \method outperforms existing methods in terms of root cause instance localization \new{(improving by 20.9\% to 368\%)} and failure type determination \new{(improving by 11.0\% to 169\%)}.

    \end{abstract}
    
}

\pagestyle{plain}

\maketitle

\section{Introduction}
\label{sec:introduction}
Microservices architecture is becoming increasingly popular for its reliability and scalability~\cite{guo2020gmta}.
Typically, it is a large-scale distributed system with dozens to thousands of service instances running on various environments (\eg physical machines, VMs, or containers)~\cite{zhou2018fault}.
Due to the complex and dynamic nature of microservice systems, the failure of one service instance can propagate to other service instances, resulting in user dissatisfaction and financial losses for the service provider.
For example, Amazon Web Service (AWS) suffered a failure in December 2021 that impacted the whole networking system and took nearly seven hours to diagnose and mitigate~\cite{aws2021summary}.
Therefore, it is crucial to timely and accurately diagnose failures in microservice systems.

\begin{figure}[!h]
    \vspacefigup
    \centering
    \resizebox{.8\linewidth}{!}{
    \includegraphics[]{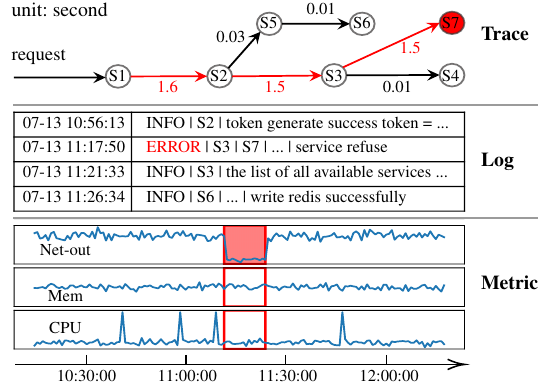}}
	\caption{
	    Multimodal data of a microservice system. 
        \new{S1 - S7 are different microservices.}
	}
	\label{fig:intro}
    \vspace{-0.6cm}
\end{figure}

To effectively diagnose failures, microservice system operators typically collect three types of monitoring data: traces, logs, and metrics.
Traces are tree-structured data that record the detailed invocation flow of user requests.
Logs are semi-structured text that record\new{s} hardware and software events of a service instance, including business events, state changes, hardware errors, \etc.
Metrics are time series indicating service status, including system metrics (\eg CPU utilization, memory utilization) and user-perceived metrics (\eg average response time, error rate).
From now on, we use the term \textit{modality} to describe a particular data type.
\new{\figref{fig:intro} shows an example of the three modalities of a microservice system.}

Automatic failure diagnosis of microservice systems has been a topic of great interest over the years, particularly when identifying the root cause instance and determining the failure type.
Most approaches rely on \textit{single-modal} data, such as traces~\cite{li2021tracerca, jin2021microhecl, yu2021microrank, guo2020gmta}, logs~\cite{lin2016logcluster, yuan2019cloud19, zhang2021onion, du2017deeplog}, or metrics~\cite{lin2018microscope, ma2020ms-rank, ma2020automap, pan2021dycause}, to capture failure patterns.
\new{However, relying solely on single-modal data for diagnosing failures is not effective enough for two reasons.}
First, a failure can impact multiple aspects of a service instance, causing more than one modality to exhibit abnormal patterns.
Using just one data source cannot fully capture these patterns and accurately distinguish between different types of failures.
Second, some types of failures may not be reflected in certain modalities, making it difficult for methods relying on that modality to identify these failures.

\new{Moreover, we conduct an empirical study on an open-source dataset to verify the necessity of combining \textit{multimodal} data for robust failure diagnosis.}
\new{As listed in \tabref{table:background}, the dataset contains failures caused by various reasons: high memory usage, incorrect deallocation, code bug, misconfiguration, network interruption, \etc.}
\new{We examine hundreds of service instance failures and conclude that combining traces, logs, and metrics (\textit{multimodal}) is crucial for accurate diagnosis.}
\new{For example, the microservice shown in \figref{fig:intro} is experiencing a failure due to missing files.}
It generated error messages in logs and a significant increase in status code 500 in related traces.
Additionally, one of its metrics, network out bytes, dropped dramatically during this failure.

These observations highlight the importance of incorporating multimodal data for robust failure diagnosis.
\new{However, combining multimodal data for diagnosing failures in microservice systems faces two major challenges:}

\begin{enumerate}
    \item \textbf{Representation of multimodal data.}
\new{The formats of metrics, logs, and traces are significantly different from each other.}
Service instance metrics are often in the form of time series (the bottom of \figref{fig:intro}), while logs are usually semi-structured text (the middle of \figref{fig:intro}) and traces often take the form of tree structures with spans as nodes (the top of \figref{fig:intro}).
It is challenging to find a unified representation of all this multimodal data that fully utilizes complementary information from each data type.
    \item \textbf{Imbalanced failure types.}
Fault tolerance mechanisms in microservice systems often result in a high ratio of normal data to failure-related data.
Some types of failures are much rarer than others, leading to an imbalance in the ratio of different types of failures (\tabref{table:background}).
\end{enumerate}

\new{To tackle the above challenges,} we present \method, an automated failure diagnosis approach that integrates trace, log, and metric data.
To form a unified representation of the three modalities with different formats and nature\new{s}, \method combines lightweight preprocessing and representation learning, which maps data from different modalities into the same vector space.
Since the labeled failures are usually inadequate to train the representation model effectively, we propose a data augmentation mechanism, which helps \method to learn the correlation between the three modalities and failures effectively.
To further enhance the accuracy of our diagnosis, \method uses historical failure patterns to train a Graph Neural Network (GNN), capturing both spatial features and possible failure propagation paths,
which allows \method to conduct root cause instance localization and failure type determination.

\new{Our contributions are summarized as follows:}

\begin{itemize}
    
    \item
        We propose \method, a multimodal data-based approach for failure diagnosis \new{(\secref{sec:approach})}.
        \method builds a dependency graph from trace and deployment data to capture possible failure propagation paths.
        Then it applies a GNN to achieve \new{a} two-fold failure diagnosis, \ie root cause instance localization and failure type determination.
        To the best of our knowledge, we are among the first to learn a unified representation of the three modalities for the failure diagnosis of microservice systems (\ie trace, log, and metric).
       
    \item
        We leverage data augmentation to improve the quality of the learned representation, which allows \method to work with limited labeled failures and imbalanced failure types.
   
    \item
        We conduct extensive experiments on two datasets, one from an open-source platform and another from a real-world microservice system \new{(\secref{sec:evaluation})}.
        The results show that when \method is trained on 160 and 80 cases, it achieves Avg@5 of 0.75 and 0.76 on the two datasets, respectively,
        improving the accuracy of \emph{root cause instance localization} by 20.9\% to 368\%.
        \new{Moreover, \method achieves F1-score of 0.84 and 0.80, improving the accuracy of \emph{failure type determination} by 11.0\% to 169\%.}
        
\end{itemize}

 Our implementation of \method is publicly available~\footnote{https://anonymous.4open.science/r/DiagFusion-378D}.

\new{
The rest of the paper is organized as follows:
\secref{sec:background} introduces the necessary background.
\secref{sec:empirical} presents the results of an empirical study of failures in microservice systems.
\secref{sec:approach} describes the overview and detailed implementation of \method in failure diagnosis.
In \secref{sec:evaluation}, we evaluate the performance and time efficiency of \method using two datasets.
\secref{sec:discussion} discusses the technical rationale, robustness, and threats to validity.
\secref{sec:related-work} presents the related work in failure diagnosis.
\secref{sec:conclusion} concludes the paper.}

\section{Background}
\label{sec:background}
\subsection{Microservice Systems and Multimodal Data}
\label{subsec:multimodal-data}

\begin{table*}[t]
    \centering
    \caption{
        Detailed information of the failures in the empirical study
    }
    \resizebox{\textwidth}{!}{%
       \begin{tabular}{llllll}
        \toprule
        \new{Failure Type} & Root Cause & Metric                                   & Log                                                                                                                                        & Trace                        & \# Failures \\
        \midrule
         \new{System stuck} & \FTone       & memory\_usage\_pct $\uparrow$            & -                                                                                                                                          & -                            & 505         \\
        \new{Process crash} & \new{Incorrect deallocation}       & memory\_stats\_active\_anon $\downarrow$ & -                                                                                                                                          & -                            & 16          \\
         \new{\FTthree} & \new{Network interruption}     & -                                        & \makecell[l]{\colorbox[HTML] {F58D6E}{ERROR} | 0.0.0.1 | \colorbox[HTML]{FFF2CC}{172.17.0.5} | M1 | uuid:\colorbox[HTML]{FFF2CC}{78fef9f0}                                              \\ information has expired, mobile phone login is invalid} & S1-\textgreater{}S2: RT=11s  & 527 \\
         \new{File missing} & \new{Code bug}      & -                                        & \makecell[l]{\colorbox[HTML] {F58D6E}{ERROR} | 0.0.0.3 | W2 | get an error [Errno 2] No such                                                                                            \\ file or directory: \colorbox[HTML]{FFF2CC}{'resources/source\_file/source\_file.csv'}} & S2-\textgreater{}S3: RT=1.5s & 36         \\
         \new{\FTfive} & \new{Misconfiguration}      & -                                        & \makecell[l]{\colorbox[HTML] {F58D6E}{ERROR} | 0.0.0.2 | B2 | \colorbox[HTML]{FFF2CC}{2768e0e0037e} | service refuse}                      & S2-\textgreater{}S4: RT=1.1s & 15          \\
        \bottomrule
    \end{tabular} 
    }
    
    \vspacetbdown
    \label{table:background}
\end{table*}

Microservice systems allow developers to independently develop and deploy functional software units (microservice).
For example, when a user tries to buy an item on an online shopping website, the user will experience item searching, item displaying, order generation, payment, \etc.
Each of these functions is served by a specific microservice.
A failure at a specific service instance can propagate to other service instances in many ways, bringing cascading failures.
However, diagnosing online failures in microservice systems is difficult due to these systems' highly complex orchestration and dynamic interaction.
To accurately find the cause of a failure, operators must carefully monitor the system and record traces, logs, and metrics.
These three modalities of monitoring data stand as the three pillars of the observability of microservice systems.
The collection and storage of instances' monitoring data are not in the scope of this paper.
The three modalities: trace, log, and metric, and their roles in failure diagnosis are described below.

\textbf{Trace.}
Traces record the execution paths of users' requests.
\figref{fig:intro} shows an example of trace at the top.
Google formally proposed the concept of traces at Dapper~\cite{sigelman2010dapper}, in which it defined the whole lifecycle of a request as a \textit{trace} and the invocation and answering of a component as a \textit{span}.
By examining traces, operators may identify microservices that have possibly gone wrong~\cite{yang2021aid, kaldor2017canopy, zhou2019mepfl, zhang2022deeptralog, li2021tracerca, li2022enjoy, yu2021microrank, liu2020traceanomaly}.
Traces can be viewed as trees, with microservices as nodes and invocations as edges.
Each subtree corresponds to a span.
Typically, traces carry information about invocations, \eg start time, caller, callee, response time, and status code.

\textbf{Log.}
Logs record comprehensive events of a service instance.
Some examples of logs are shown in the middle of \figref{fig:intro}.
Logs are generated by developers using commands like \textit{printf}, \textit{logging.debug}, \textit{logging.error}.
They provide an internal picture of a service instance.
By examining logs, operators may discover the actual cause of why an instance performs not well.
Typically, logs consist of three fields: timestamp, verbosity level, and raw message~\cite{he2017drain}.
Four commonly used verbosity levels, \ie INFO, WARN, DEBUG, and ERROR, indicate the severity of a log message.
The raw message of a log conveys detailed information about the event.
To utilize logs more effectively, researchers have proposed various parsing techniques to extract templates and parameters, \eg FT-Tree~\cite{zhang2017ft-tree}, Drain~\cite{he2017drain}, POP~\cite{he2017pop}, MoLFI~\cite{messaoudi2018molfi}, Spell~\cite{du2018spell}, and Logram~\cite{dai2020logram}.

\textbf{Metric.}
Various system-level metrics (\eg CPU utilization, memory utilization) and user-perceived metrics (\eg average response time) are configured for monitoring system instances.
Each metric is collected at a predefined interval, forming a time series, as shown at the bottom of \figref{fig:intro}.
These metrics track various aspects of performance issues.
By examining metrics, operators can determine which physical resource is anomalous or is the bottleneck~\cite{sun2021ctf, su2021detecting, shen2021time, ma2021jump, li2021multivariate, dai2021sdfvae}.

\new{In addition to trace, log, and metric, \textbf{deployment data} is also important to failure diagnosis.}
A microservice system comprises many hardware and software assets that form complicated inter-relationships.
Operators must carefully record these relationships (\aka deployment data) to keep high maintainability of the system.
\new{Leveraging deployment data enables the understanding of failure propagation paths and characteristics.}

\subsection{Preliminaries}

\textbf{Representation learning.}
Representation learning has been widely used in natural language processing tasks, usually in the form of word embedding.
Popular techniques of representation learning include static representation like word2vec~\cite{mikolov2013word2vec}, GloVe~\cite{pennington2014glove}, fastText~\cite{bojanowski2017fasttext}, and dynamic representation like ELMo~\cite{Matthew2018elmo}, BERT~\cite{devlin2018bert}, GPT~\cite{brown2020language}.
With the similarities between logs and natural languages, representation learning can be applied to extract log features~\cite{meng2019loganomaly}.
We employ fastText to learn a unified representation of events from multimodal data.
Compared to word2vec and GloVe, fastText can utilize more information~\cite{bojanowski2017fasttext}.
\new{We employ fastText to learn a unified representation of the multimodal data.}

\new{In essence, fastText is a neural network model that processes words as input and takes the output from the hidden layer (a vector of real numbers) as its representation. It can be trained in both supervised and unsupervised modes, but the supervised mode generally yields more accurate results due to its incorporation of label information. In the supervised training mode, the neural network is optimized by predicting the class of the document. Once the training is completed, fastText can be used to provide vectorized representations (\ie embeddings) for any given input.}

\textbf{Graph neural network.}
GNN can effectively model data from non-euclidean space, thereby being popular among fields with graph structures, \eg social networks, biology, and recommendation systems.
Popular GNN architecture includes Graph Convolution Network (GCN)~\cite{kipf2016gcn}, GraphSAGE~\cite{hamilton2017GraphSAGE}, Graph Attention Network (GAT)~\cite{zhou2022hybrid}, \etc.
GNNs apply graph convolutions, allowing nodes to utilize their information and learn from their neighbors through message passing.
There are numerous components in microservice systems that interconnect with each other.
Thus graph structure is suitable to model microservice systems, and we employ GNN to learn the propagation patterns of historical failure cases.

\subsection{Problem Statement}
\label{content:problem-statement}

When a failure occurs, operators need to localize the root cause instance and determine what has happened to it to achieve timely failure mitigation.
For large-scale microservice systems, the first task is a ranking problem: to rank the root cause instance higher than other instances.
We use the term \textit{root cause instance localization} to name this task (Task~\#1).
The second task is a classification problem: to classify the failure into a predefined set of failure types.
We use the term \textit{failure type determination} to name this task (Task~\#2).

After each failure, operators will carefully conduct a post-failure analysis: labeling its root cause instance and its failure type.
Additionally, chaos engineering can generate a large number of failure cases~\cite{zhang2019chaos}.
It can enlarge the number of failure cases and enrich the types of failures.
We train \method based on these failure cases.

\section{Empirical Study}
\label{sec:empirical}
\label{sec:emprical_study}

Most existing failure diagnosis methods are based on single-modal data.
However, these methods cannot fully capture the patterns of failed instances, leading to ineffective failure diagnosis. 
We conduct an empirical study conducted on Generic AIOps Atlas (GAIA)\footnote{\url{https://github.com/CloudWise-OpenSource/GAIA-DataSet}} dataset to show the ineffectiveness of these methods.
        The dataset is collected from a simulation environment consisting of 10 microservices, two database services (MySQL and Redis), and five host machines.
        The system serves mobile users and PC users.
        Operators injected five types of failures, including \new{system} failures
            \new{(System stuck and Process crash)}
        and \new{service} failures
            \new{(Login failure, File missing, and Access denied)}.
        The failure injection record is provided along with the data.
\tabref{table:background} lists some typical symptoms of failures.
We can see that no modality alone can distinguish the patterns of these five types of failures.
It also shows that traces, logs, and metrics may display different anomalous patterns when a failure occurs.
Mining the correlation between multimodal data can provide operators with a more comprehensive understanding of failures.

Besides, \tabref{table:background} shows that some failures occur much more frequently than others.
For example, the total occurrences of \new{\textit{Process crash}}, \new{\textit{File missing}}, and \new{\textit{Access denied}} (67) equals only 12\% of the occurrences of \new{\textit{Login failure}} (527).

\begin{figure}[!h]
    \vspacefigup
    \centering
    \includegraphics[]{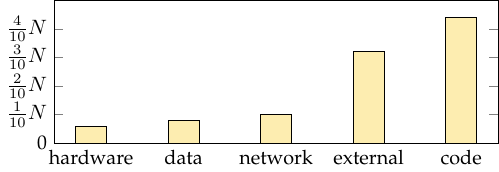}
    \caption{
        The distribution of failure types at a large-scale real-world microservice system.
    }
    \label{fig:empirical-real-world}
    \vspacefigdown
    \vspace{-8pt}
\end{figure}

To further understand the distribution of failure types in the production environment, we investigated $N$ failures in a microservice system of Microsoft.
Due to the company policy, we have to hide some details of these failures.
The failures of the studied system are recorded in the Incident Management System (IcM) of Microsoft, where a failure is centralized handled, including the detection, discussion, mitigation, and post-failure analysis of failures.
The IcM data of failures are persistently stored in a database.
We query the failure records from the database within the time range from 2021 August to 2022 August.
We only keep the failures with the status of ``completed'', for their post-failure analyses have been reviewed.
In the \textit{root cause} field of post-failure analysis, operators categorize the failures into the following types: code, data, network, hardware, and external.
We can see from \figref{fig:empirical-real-world} that different failure types are imbalanced regarding the number of failure cases.
The imbalanced data poses a significant challenge because most machine learning methods perform poorly on failure types with fewer occurrences. 

\section{Approach}
\label{sec:approach}
\subsection{Design Overview}

\begin{figure}[!h]
    \vspacefigup
    \centering
    \includegraphics[]{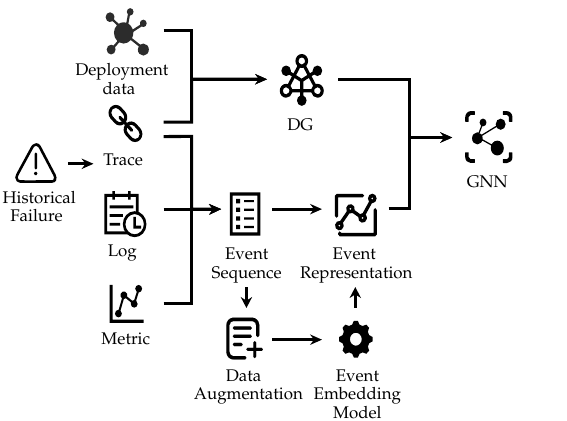}
    \caption{
        The training framework of \method.
    }
    \label{fig:offline}
    \vspacefigdown
\end{figure}

In this paper, we propose \method, which combines the modality of trace, log, and metric for accurate failure diagnosis.
The training framework of \method is summarized in \figref{fig:offline}.
First, \method extracts events from raw traces, logs, and metrics data and serializes them by their timestamps.
Then, we train a neural network to learn the distributed representation of events by encoding events into vectors.
The challenge of data imbalance is overcome through data augmentation during model training.
We unify three modalities with different natures by turning unstructured raw data into structured events and vectors.
Then we combine traces with deployment data to build a dependency graph (DG) of the microservice system.
After that, the representations of events and \kgshort are glued together by \new{a} GNN.
We train GNN using historical failures to learn the propagation pattern of system failures.

After the training stage, we save the event embedding model and the GNN.
\figref{fig:realtime} depicts the real-time failure diagnosis framework of \method.
The trigger of \method can be alerts generated through predefined rules.
When a new failure is alerted, \method will perform a real-time diagnosis and give the results back to operators.   

\subsection{Unified Event Representation}
\label{sec:event-representation}

\label{content:event-extraction}

\method unifies the three modalities by extracting events from the raw data and encoding them into vectors.
Specifically, it collects failure-indicative events by leveraging effective and lightweight methods, including anomaly detection techniques for metrics and traces and \new{template} parsing techniques for logs.
Then, it trains a fastText~\cite{bojanowski2017fasttext} model on event sequences to generate embedding vectors of events.

\new{First, we introduce the \textit{instances} in a microservice system.}
Microservice systems have the advantage of dynamic deployment by utilizing the container technique. In this paper, we use the term \textit{instance} to describe a running container and the term \textit{service group} to describe the logical component that an instance belongs to. For example, \textit{Billing} is a service group in a microservice system, and \textit{Billing\_cff19b} denotes an instance, where \textit{cff19b} is the container id.
\new{Below we will describe the event extraction from different modalities.}

\textbf{Trace event extraction.}
Traces record calling relationships between services.
We group trace data by its caller and callee services.
\method will examine multiple fields inside a trace group.
Under different implementations of trace recording, trace data can carry different fields, \eg response time and status code, which reflect different aspects of operators' interest\new{s}.
We apply an anomaly detection algorithm (\ie 3-sigma) for numerical fields like response time to detect anomalous behaviors.
For categorical fields like status code, we count the number of occurrences of each value.
If the count of some value increases dramatically, we determine that this field is anomalous.
We determine that a group of caller and callee is anomalous if one of its fields becomes anomalous.
The extracted trace events are in the form of tuple \textit{<timestamp, caller-instance-id, callee-instance-id>}.

\textbf{Log event extraction.}
Logs record detailed activities of an instance (service or machine).
We perform log parsing for log event extraction using Drain~\cite{he2017drain}, which has been proven to be effective in practice.
Drain uses a fixed depth parse tree to distinguish the template part and the variable part of log messages.
For example, in the log message ``uuid: 8fef9f0 information has expired, mobile phone login is invalid'', ``uuid: ****** information has expired, mobile phone login is invalid'' is the template part\new{,} and ``8fef9f0'' is the variable part.
After we get the template part of a log message, we hash the string of the template part to obtain an event template id.
The extracted log events are in the form of tuple \textit{<timestamp, instance-id, event-template-id>}.

\textbf{Metric event extraction.}
Metrics are also recorded at the instance level.
We perform 3-sigma to detect anomalous metrics.
When the value of a metric exceeds the upper bound of 3-sigma, the anomaly direction is \textit{up}.
Similarly, the anomaly direction is \textit{down} if the value is below the lower bound.
The extracted metric events are in the form of tuple \textit{<timestamp, instance-id, metric-name, anomaly-direction>}.

\begin{figure}[t]
    \vspacefigup
    \centering
    \includegraphics[]{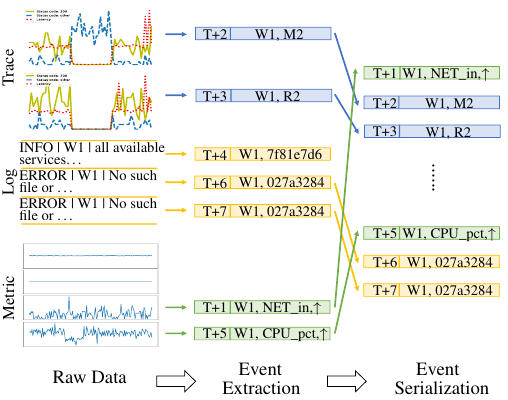}
    \caption{
        The event extraction and serialization process using traces, logs, and metrics.
    }
    \label{fig:events}
    \vspacefigdown
    \vspace{-5pt}
\end{figure}

\new{The above extraction provides events from different modalities.}
Despite the differences in \new{raw} data, all extracted events share two fields, namely \textit{timestamp} and \textit{instance-id}.
These are the keys to unifying different modalities.
We group events by \textit{instance-id} and serialize events in the same group by \textit{timestamp}.
\figref{fig:events} shows the event extraction and serialization process for one instance.
The event sequence of instance $i$ is denoted by $E_{i}$.

\new{After getting the event sequence of every instance, we further assign labels to every event sequence according to operators' post-failure analysis.}
\new{Original} failure labels are often in the form of tuple \textit{<root cause instance-id, failure type>}.
To fully utilize the label information, we relabel event sequences in an instance-wise manner.
Specifically, the root cause instance's event sequence is labeled by the actual failure type, while other instances' event sequences are labeled as ``non-root-cause''.
A microservice system with $p$ historical failures and $q$ instances results in $N=p\times q$ event sequences after relabeling.
Then, we learn unified representations from these relabeled historical event sequences using the event embedding model.

\new{With event sequence and instance labeling, we can transform events into vectors.}
\new{We use the term \textit{event embedding} to describe the mapping of events to real number vectors.}
Specifically, we train a fastText model on the event sequences to obtain the vectorized representation for events from all three modalities.
FastText is a neural network originally proposed for text classification.
For a document with word sequences, fastText extracts $n$-grams from it and predicts its label.
In our scenario, we replace word sequences with event sequences and replace document labels with failure types.
The training of fastText minimizes the negative log-likelihood over classes:
\begin{equation}
    \min_{f}-\frac{1}{N}\sum_{n=1}^{N}y_{n}\log\left(f\left(x_{n}\right)\right)
\end{equation}
where $x_{n}$ is the normalized bag of features of the $n$-th event sequence, $y_{n}$ denotes the relabeled information, and $f$ is the neural network.
We treat fastText's output as the vectorized representation of events.
The training detail of the event embedding model is described in \secref{subsec:training}.

\subsection{Graph Neural Network}
\label{sec:GNN}
\label{content:GNN}

In the event representation process, \method captures the local features of instances.
However, failures can propagate between instances, so we need to have a global picture of the system, \ie how a failure will affect the system.
To this end, we employ \new{a} GNN to learn the failure propagation between service instances and integrate all the information of the whole system.

\new{To leverage a GNN, it is essential to consider both \textit{nodes} and \textit{edges} within a graph.
The \textit{nodes} in a GNN corresponds to the instances in a microservice system.}
An instance is characterized by its anomalous events in \method.
We represent an instance $i$ by averaging all of its events:
\begin{equation}
    h_{i}^{\left(0\right)}=\frac{1}{\lvert E_{i}\rvert}\sum_{\forall e\in E_{i}}{\mathcal{V}_1(e)}
\end{equation}
where $E_{i}$ is the extracted event sequences, and ${\mathcal{V}_1(e)}$ is the vectorized representation of event $e$ learned by the event embedding model.

\begin{figure}[!htb]
    \vspacefigup
    \centering
    \includegraphics[]{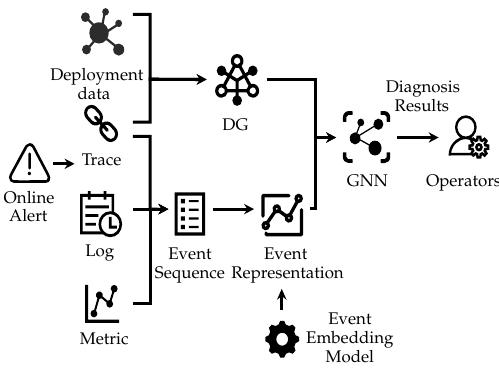}
    \vspace{-10pt}
	\caption{
	    Real-time failure diagnosis.
	}
	\label{fig:realtime}
\end{figure}

\new{The \textit{edges} in a GNN correspond to the dependency graph in a microservice system.}
There are two dominant ways of propagation failure between services: function calling or resource contention~\cite{outagescope21wang}.
So we combine traces and deployment data to capture probable failure propagation paths.
Specifically, we aggregate traces to get a call graph.
Then we add two directed edges for each pair of caller and callee, with one pointing from the caller to the callee and the other in the reverse direction.
\new{From deployment data, we add edges between two instances if they are co-deployed, \ie sharing resources.}

After obtaining the dependency graph and instance representations, we \new{employ} GNN to learn the failure propagation pattern by its message-passing mechanism.
At the $K$-th layer of GNN, we apply topology adaptive graph convolution~\cite{zhou2022graph} and update the internal data of instances according to:
\vspace{-3pt}
\begin{equation}
    H^{K}=\sum_{k=0}^{K}\left(D^{-1/2}AD^{-1/2}\right)^{k}X\Theta_{k}
\end{equation}
where $A$ denotes the adjacency matrix, $D_{ii}=\sum_{j=0}A_{ij}$ is a diagonal degree matrix, $\Theta_{k}$ denotes the linear weights to sum the results of different hops together.

\new{Finally, } we add a MaxPooling layer \new{as the readout layer} to integrate the information of the whole microservice system.
Following the MaxPooling layer, there is a fully connected layer where each neuron corresponds to either a service group with possible root cause instances for task~\#1 or a failure type for task~\#2.

\subsection{Training of \method}
\label{subsec:training}

\method applies a two-phase training strategy to learn the failure pattern of a microservice system. First, it trained the event embedding model with data augmentation. Then it trains the GNN with a joint learning technique.

\subsubsection{Training of Event Embedding Model}

\method employs a data augmentation strategy to enrich the training dataset and reduce the model's bias towards the majority class.
First, we train our event embedding model on the original data.
The trained neural network, denoted by $f_0$, maps events to the vector space $\mathcal{V}_0$.
To increase the number of failure cases, we add new event sequences for each failure type (including ``non-root-cause'') by randomly taking an event sequence of that type and replacing one of the events with its closest neighbor (determined by Euclidean distance) in $\mathcal{V}_0$.
After all failure types are expanded to a relatively large size, \eg 1000, we can obtain a more balanced training set.
Further details on the choice of the expanding size can be found at \secref{content:hyperparameter}.
Then we train the event embedding model again ($f_1$) on the expanded data and regard the representations generated in this round ($\mathcal{V}_1$) as the final unified event representations.

\subsubsection{Training of Graph Neural Network}

\begin{figure}[!t]
    \vspacefigup
    \centering
    {\includegraphics[width=250.52pt,height=149.64pt]{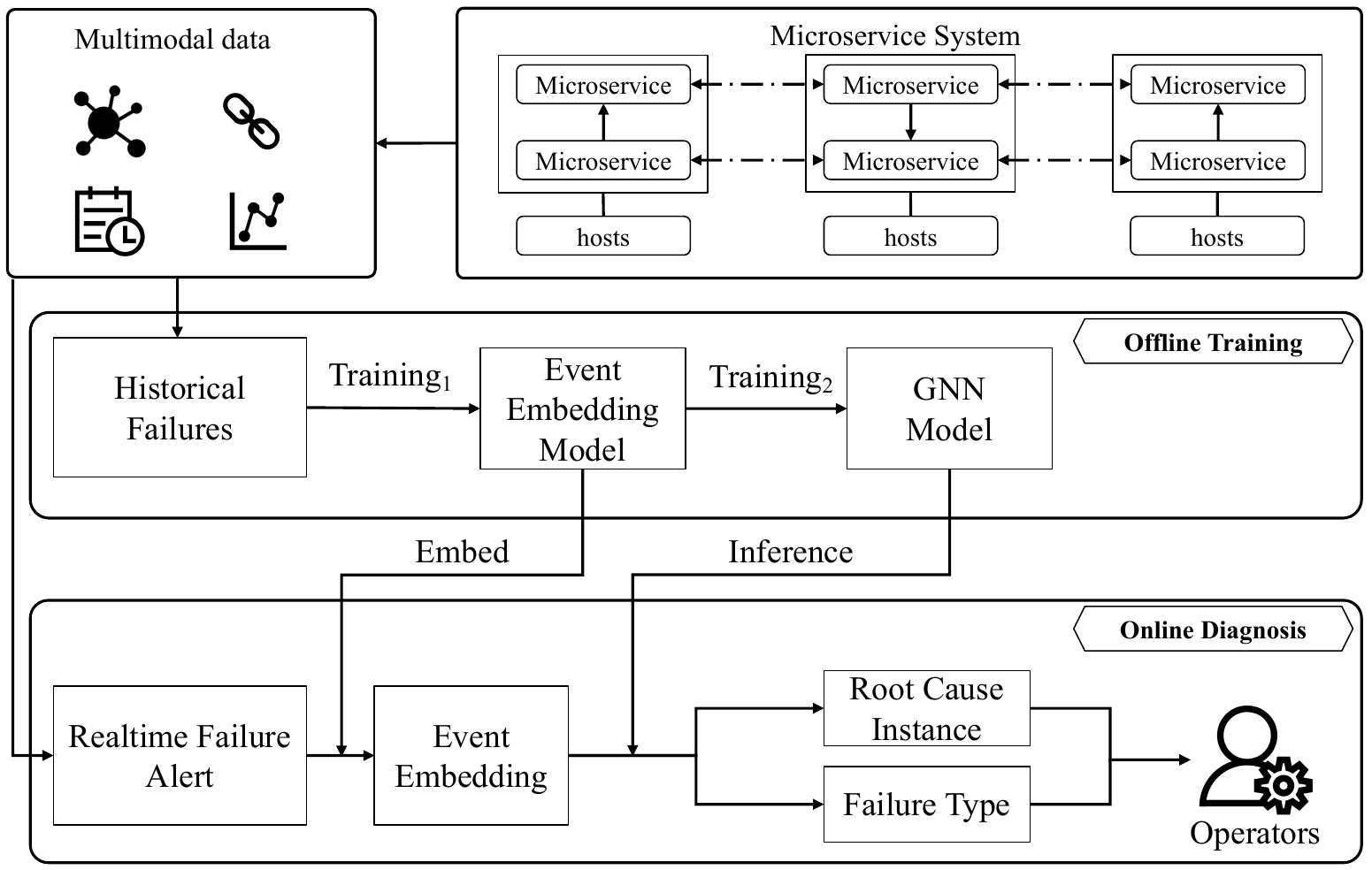}}
	\caption{
	    \new{Integration of \method with a microservice system.}
	}
	\label{fig:overall}
\end{figure}

\newlength{\casestudylength}
\setlength{\casestudylength}{\textwidth}
\begin{figure*}[t]
    \vspacefigup
	\centering
	\subfigure[]{ 
		\centering
        \includegraphics[]{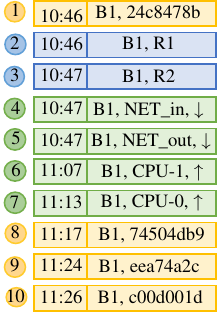}
		\label{fig:case-study-b}
	}
    \hspace{-10pt}
	\subfigure[]{
		\centering
        \includegraphics[]{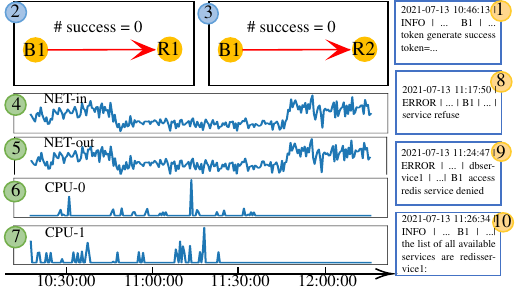}
		\label{fig:case-study-c}
	}
    \hspace{-10pt}
	\subfigure[]{
		\centering
        \includegraphics[]{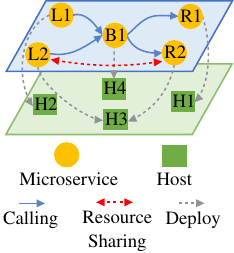}
		\label{fig:case-study-a}
	}
	\vspacefigup
    \caption{
        A running example of \method.
        (a): the serialized multimodal event sequence of the root cause instance (B1);
        (b): the original data corresponding to the event sequence;
        (c): part of the dependency graph in this failure.
    }
	\label{fig:case-study}
\end{figure*}

\new{
We train the GNN in a joint learning fashion to fully utilize the shared information between tasks \#1 and \#2.
Then we combine the trained GNN with a ranking strategy to better fit the nature of microservice systems.}

\new{\textbf{Ranking strategy}.}
One of the advantages of microservice systems is that the architecture allows dynamic deployment of service instances.
Thus, service instances are constantly being created and destroyed.
However, when it comes to failure diagnosis, this kind of flexibility raises a challenge for learning-based methods.
The failure diagnosis model will have to be retrained frequently if the output layer directly outputs the probability of being the root cause instance for each instance since many instances can be created or destroyed after the model training is finished.
We add an extract step in \method to overcome this challenge.
Instead of directly determining the root cause instance, \method is trained on service groups, the logical aggregation of service instances, for task~\#1.
Then \method ranks the instances inside a candidate service group by the length of their event sequences.
The instance with more anomaly events will be ranked higher and likely be the root cause instance.

\textbf{Joint learning}.
Intuitively, the two tasks of failure diagnosis, \ie root cause instance localization and failure type determination, share some knowledge in common.
For a given failure, the only difference between task~\#1 and task~\#2 lies in their labels.
So \method integrates a joint learning mechanism to utilize the shared knowledge and reduce the training time.
(Training two models separately requires twice the time otherwise.)
Specifically, the joint loss function is:
\begin{equation}
    - \frac{1}{F} \sum_{i=1}^{F}
    \left(
        \sum_{j = 1}^{S} y(s)_{i, j} \log{ p(s)_{i, j} }
        +
        \sum_{k = 1}^{T} y(t)_{i, k} \log{ p(t)_{i, k} }
    \right)
\end{equation}
where $F$ is the number of historical failures, $S$ is the number of service groups, $T$ is the number of failure types, $y\left(s\right)$ is the root cause service group labeled by operators, $y\left(t\right)$ is the failure type, $p\left(s\right)$ is the predicted service group, and $p\left(t\right)$ is the predicted failure type.

\subsection{Real-time failure diagnosis}
After the training stage, we save the trained event embedding model and the GNN.
When a new failure is alerted, \method performs a real-time diagnosis process as shown in \figref{fig:realtime}.

\subsubsection{Running Example}

\new{\figref{fig:overall} shows how \method can be integrated with microservice systems.}
To \new{better explain} how \method diagnoses failure, we demonstrate the workflow of \method using one real-world failure from \Done.
At 10:46, service instance B1 encounters a failure of access denied.
\figref{fig:case-study} shows the original data, event sequence, and the \kgshort.
From \figref{fig:case-study-b}, we can see that failure-indicative events from different modalities are temporally intertwined.
\new{Then the GNN predicts service group ``B'' and failure type ``access denied''. Further ranking within the service group ``B'' gives ``B1'' as the Top1 instance. The overall process takes less than 10 seconds. Thus, \method effectively addresses tasks \#1 and \#2.}

\section{Evaluation}
\label{sec:evaluation}

In this section, we evaluate the performance of \method using two real-world datasets.
We aim to answer the following research questions (RQs):

\noindent
\textbf{RQ1}:
How effective is \method in failure diagnosis?

\noindent
\textbf{RQ2}:
Does each component of \method have significant contributions to \method's performance?

\noindent
\textbf{RQ3}:
Is the computational efficiency of \method sufficient for failure diagnosis in the real world?

\noindent 
\textbf{RQ4}:
What is the impact of different hyperparameter\new{s}?

\subsection{Experimental Setup}
\label{sec:eval-exp-setup}

\begin{figure*}[!t]
    \vspacefigup
    \vspacefigup
\begin{minipage}{\textwidth}
\newlength{\downstreamlength}
\setlength{\downstreamlength}{4.7cm}

\begin{center}
\includegraphics[]{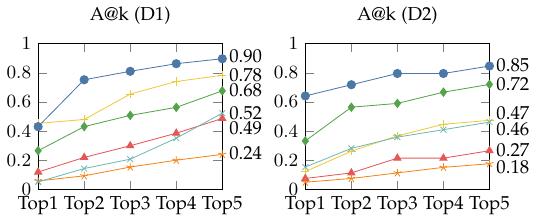}
\includegraphics[]{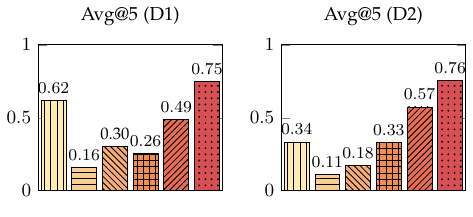}
\includegraphics[]{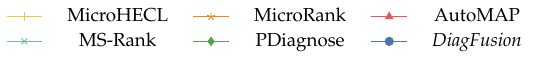}
\includegraphics[]{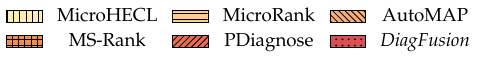}
\end{center}

\vspace{-7pt}

\label{fig:exp_2}
\end{minipage}
	\vspacefigup
    \caption{
        Effectiveness of root cause instance localization (Task \#1).
    }
	\label{fig:effect-instance}
	\vspacefigdown
\end{figure*}

\subsubsection{Dataset}
To evaluate the performance of \method, we conduct extensive experiments on two datasets collected from two microservice systems under different business backgrounds and architectures, \Done and \Dtwo.
To prevent data leakage, we split the data of \Done and \Dtwo into training and testing sets according to their start time, \ie we use data from the earlier time as the training set and data from the later time as the test set.
Detailed information is listed in \tabref{table:datasets}.
The systems that produce \Done and \Dtwo are as follows:
\begin{enumerate}[leftmargin=*]
    \item \Done.
        The details of \Done \new{are} elaborated in Section~\ref{sec:emprical_study}.
    \item \Dtwo.
        The second dataset is collected from the management system of a top-tier commercial bank.
        The studied system consists of 14 instances, including microservices, web servers, application servers, databases, and dockers.
        Due to the non-disclosure agreement, we cannot make this dataset publicly available.
        Two experienced operators examined the failure records from January 2021 to June 2021.
        They classified the failures into five types of failures, \ie CPU-related failures, memory-related failures, JVM-CPU-related failures, JVM-memory-related failures, and IO-related failures.
        The classification was done separately, and they checked the labeling with each other to reach a consensus.
\end{enumerate}

\begin{table}[t]
\vspacetbup
\caption{
    Detailed information of datasets
}
\vspacetbup
\resizebox{\linewidth}{!}{
\begin{tabular}{c|cccll}
\toprule
Dataset             & \# Instances         & \# Training           & \# Test               & \multicolumn{2}{l}{\# Records} \\ \midrule
\multirow{3}{*}{\Done} & \multirow{3}{*}{17} & \multirow{3}{*}{160} & \multirow{3}{*}{939} & trace       & 2,321,280       \\
                    &                     &                      &                      & log         & 87,974,577      \\
                    &                     &                      &                      & metric      & 56,684,196      \\ \midrule
\multirow{3}{*}{\Dtwo} & \multirow{3}{*}{18} & \multirow{3}{*}{80} & \multirow{3}{*}{79}  & trace       & 1,123,200       \\
                    &                     &                      &                      & log         & 21,356,923      \\
                    &                     &                      &                      & metric      & 8,228,010     
                    \\ \bottomrule
\end{tabular}}
\label{table:datasets}
\vspacetbdown
\end{table}

\subsubsection{Baseline Methods}

We select six advanced single-modal-based methods (two for trace (\ie MicroHECL \cite{jin2021microhecl}, MicroRank \cite{yu2021microrank}), two for log (\ie Cloud19 \cite{yuan2019cloud19}, LogCluster \cite{lin2016logcluster}), and two for metric (\ie AutoMAP \cite{ma2020automap}, MS-Rank \cite{ma2020ms-rank})), and two multimodal-based methods (\ie PDiagnose \cite{hou2021pdiagnose}, CloudRCA \cite{zhang2021cloudrca}) as the baseline methods.
More details can be found in \secref{sec:related-work}.
Among the baseline methods, Cloud19, LogCluster, and CloudRCA cannot address Task \#1 (root cause instance localization), while MicroHECL, MicroRank, AutoMAP, MS-Rank, and PDiagnose cannot address Task \#2 (failure type determination).
Therefore, we divide the baseline methods into two groups to evaluate the performance of Task \#1 and Task \#2, respectively: MicroHECL, MicroRank, AutoMAP, MS-Rank, and PDiagnose for Task \#1, Cloud19, LogCluster, and CloudRCA for Task \#2.

We configure the parameters of all these methods according to their papers. Specifically, we use the same configuration for parameter settings explicitly mentioned in the papers and not limited to a particular dataset (e.g., significance level, feature dimension).  For parameter settings that apply to a particular dataset (e.g., window length, period), we adapt them according to the range the papers provide or to our data.

\subsubsection{Evaluation Metrics}
As stated in \secref{content:problem-statement}, \method aims to localize the root cause instance and determine the failure type. We carefully select different evaluation metrics for both tasks to better reflect the real-world performance of all selected methods.

For Task \#1, we use \textit{Top-k accuracy} (A@k) and \textit{Top-5 average accuracy} (Avg@5) as the evaluation metrics.
A@k \new{is a well-adopted metric that} quantifies the probability that top-k instances output by each method indeed contain the root cause instance\new{\cite{jin2021microhecl}}.
Formally, given $|A|$ as the test set of failures, $RC_i$ as the ground truth root cause instance, $RC_s \left[k\right]$ as the top-k root cause instances set generated by a method, A@k is defined as:
\begin{equation}
    A@k=
        \frac{1}{|A|} 
        \sum_{a \in A}
        \begin{cases}
            1, & \text { if } {RC_i}_a  \in {RC_s}_a \left[k\right] \\ 
            0, & \text { otherwise }
        \end{cases}
\end{equation}
Avg@5 \new{is another popular metric that} evaluates a method's overall capability of localizing the root cause instance\new{\cite{meng2020microcause}}. 
In practice, operators often examine the top 5 results.
Avg@5 is calculated by:
\begin{equation}
    A v g @ 5=\frac{1}{5} \sum_{1 \leq k \leq 5} A @ k
\end{equation}

For Task \#2, which is a multi-class classification problem, 
we use the weighted average \textit{precision}, \textit{recall}, and \textit{F1-score} to test the performances.
\new{These metrics have been selected based on a previous study~\cite{ma2020isquad} as a reliable way to assess the model's effectiveness in this specific context.}
With True Positives (TP), False Positives (FP), and False Negatives (FN), the calculation is given by $\text{F1-score}=2\times\frac{precision\times recall}{precision+recall}$, where $precision=\frac{\mathrm{TP}}{\mathrm{TP}+\mathrm{FP}}$ and $recall=\frac{\mathrm{TP}}{\mathrm{TP}+\mathrm{FN}}$.

\begin{table}[t]
    \centering
    \caption{
        Effectiveness of failure type determination (Task \#2)
    }
    \label{table:eval-effect-type}
    \vspacetbup
    \resizebox{\linewidth}{!}{
    \begin{tabular}{c|cccccc}
        \toprule
        \multirow{2}[2]{*}{Method} & \multicolumn{3}{c}{\Done} & \multicolumn{3}{c}{\Dtwo} \\
        \cmidrule(lr){2-4} \cmidrule(l){5-7} 
        & Precision & Recall & F1-score & Precision & Recall & F1-score  \\
        \midrule
        \method                 & \textbf{0.860} &  \textbf{0.829} & \textbf{0.839} & \textbf{0.822} & \textbf{0.797} & \textbf{0.800}  \\
        Cloud19                 & 0.774          & 0.774          & 0.756          & 0.526          & 0.278          & 0.297                    \\
        LogCluster              & 0.615          & 0.477          & 0.336          & 0.521           & 0.722         & 0.605                    \\
        CloudRCA                & 0.436          & 0.453          & 0.357          & 0.589          & 0.506          & 0.538                  \\
        \bottomrule
    \end{tabular}}
    \vspacetbdown
    \vspace{-5pt}
\end{table}

\subsubsection{Implementation}
We implement \method and baselines with Python 3.7.13, PyTorch 1.10.0, scikit-learn 1.0.2, fastText 0.9.2, and DGL 0.9.0. 
We run the experiments on a server with 12 $\times$ Intel(R) Xeon(R) CPU E5-2650 v4 @ 2.20GHz and 128G RAM (without GPUs).
We repeat every experiment five times and take the average result to reduce the effect of randomness.

\begin{table*}[t]
    \begin{minipage}{0.61\textwidth}
    \centering
    \caption{
        Contributions of components
    }
    
    \begin{tabular}{cc|ccccccc}
        \toprule 

        & \multirow{2}[2]{*}{Method} 
        & \multicolumn{4}{c}{Task \#1} 
        & \multicolumn{3}{c}{Task \#2}\\
        \cmidrule(lr){3-6} 
        \cmidrule(l){7-9}
        & 
        & A@1 & A@3 & A@5 & Avg@5 
        & Precision & Recall & F1-score\\
        \midrule
        \multirow{6}{*}{\Done}
        & \method 
        & \textbf{0.419} & \textbf{0.813} & \textbf{0.914} & \textbf{0.750} 
        & \textbf{0.860} & \textbf{0.829} & \textbf{0.839} \\
        & C1
        & 0.341          & 0.678          & 0.833          & 0.641
        & 0.809          & 0.793          & 0.779          \\
        & C2                         
        & 0.306          & 0.639          & 0.780          & 0.594 
        & 0.780          & 0.765          & 0.768          \\
        & C3
        & 0.309          & 0.632          & 0.770          & 0.588           
        & 0.773          & 0.797          & 0.781          \\
        & C4                       
        & 0.359          & 0.657          & 0.760          & 0.616          
        & 0.351          & 0.102          & 0.104          \\
        & C5                  
        & 0.419          & 0.809          & 0.905          & 0.744          
        & 0.089          & 0.102          & 0.095          \\
        \midrule
        \multirow{6}{*}{\Dtwo}
        & \method                    
        & 0.646          & \textbf{0.848} & \textbf{0.873} & \textbf{0.790} 
        & \textbf{0.822} & \textbf{0.797} & \textbf{0.800} \\
        & C1             
        & 0.304          & 0.506          & 0.646          & 0.471          
        & 0.567          & 0.608          & 0.576          \\
        & C2
        & 0.646          & 0.823          & 0.861          & 0.780          
        & 0.793          & 0.734          & 0.753          \\
        & C3                         
        & \textbf{0.671}          & 0.823          & 0.848          & 0.785          
        & 0.787          & 0.747          & 0.747          \\
        & C4                        
        & 0.494          & 0.620          & 0.646          & 0.587      
        & 0.780          & 0.595          & 0.639          \\
        & C5                    
        & 0.582          & 0.709          & 0.709          & 0.671      
        & 0.778          & 0.797          & 0.764          \\
        \bottomrule
    \end{tabular}
    \label{table:contributions-of-components}
    \vspacetbup
    \end{minipage}\quad
    \begin{minipage}{0.37\textwidth}
        \caption{
            The comparison of training time (Offline) and diagnosis time  (Online) per case (``-'' means no need training)
        }
        \vspace{14pt}
    \vspacetbup
    \begin{tabular}{c|cccc}
    \toprule
    \multirow{2}[2]{*}{Method}    & \multicolumn{2}{c}{\Done} & \multicolumn{2}{c}{\Dtwo} \\
        \cmidrule(lr){2-3} \cmidrule(l){4-5}
        & Offline    & Online            & Offline    & Online            \\
      \midrule
    \method     & 11.02 & 10.95 & 3.59 & 3.26 \\
    MicroHECL  & - & 65.98 & -  & 28.40             \\
    MicroRank  & 22.9    & 34.47 & 53.2   & 54.94    \\
    Cloud19    & 0.41     & 0.03   & 0.03   & 0.03   \\
    LogCluster & \textless{}0.1   & \textless{}0.01   & 0.2   & \textless{}0.01   \\
    AutoMap    & -  & 0.299  & -          & 0.511            \\
    MS-Rank    & -  & 1.14   & -           & 12.94            \\
    PDiagnose  & -  & 42.51   & -          & 68.74          \\ 
    CloudRCA  & 1.43  & 0.06   & 0.83          & 0.07          \\ 
    \bottomrule
    \end{tabular}
    \label{table:efficiency}
    \vspacetbup
    \end{minipage}\quad
    \vspace{-5pt}
\end{table*}

\subsection{Overall Performance (RQ1)}
\label{content:eval-effect}

To demonstrate the effectiveness of \method, we compare it with the baseline methods on Task \#1 and Task \#2.

The comparison result of Task \#1 is shown in \figref{fig:effect-instance}.
\method achieves the best performance. 
Specifically, the A@1 to A@5 of \method are almost the best on \Done and \Dtwo.
More specifically, the Avg@5 of \method exceeds 0.75 on both \Done and \Dtwo, respectively.
It is at least 0.13 higher on both datasets than baselines using single-modal data due to the advantage of using multimodal data. 
Compared with PDiagnose, which also uses multimodal data, the Avg@5 of \method is higher by at least 0.18. 
This indicates that learning from historical failures improves the accuracy of diagnosis significantly.

The result of Task \#2 is shown in \tabref{table:eval-effect-type}.
For this task, \method is better than almost all baselines.
On \Done, the precision, recall, and F1-score of \method are over 0.80.
On \Dtwo, \method manages to maintain an F1-score of 0.80, which is at least 0.195 higher than the baselines.
\new{Considering both systems and tasks, DiagFusion consistently demonstrates superior performance, thereby substantiating its effectiveness.}

\subsection{Ablation Study (RQ2)}
\label{subsec:rq2}

To evaluate the effects of the three key technique contributions of \method: 
\begin{enumerate*}[1)]
    \item data augmentation;
    \item fastText embedding;
    \item \kgshort and GNN
\end{enumerate*},
we create five variants of \method.
\begin{enumerate*}[\textbf{C\arabic*:}]
    \item Remove the data augmentation.
    \item Use word2vec embedding instead of fastText.
    \item Use GloVe embedding instead of fastText.
    \item Replace the GNN output layer with a decision tree.
    \item Replace the GNN output layer with a kNN model.
\end{enumerate*}

\tabref{table:contributions-of-components} lists that \method outperforms all the variants on \Done and \Dtwo, demonstrating each component's significance. 
When removing the data augmentation (\textbf{C1}), the performance reduces across the board as models trained from imbalanced data are more likely to bias predictions toward classes with more samples. Data augmentation can alleviate this problem.
The performance becomes worse when replacing fastText
embedding strategy (\textbf{C2} \& \textbf{C3}). 
\new{The reason is that fastText can learn from operators' failure labeling as well as co-occur relationships, while word2vec and GloVe can only learn from the co-occur relationships between events.}
Replacing the GNN output layer with classifiers such as decision trees and kNN (\textbf{C4 \& C5}) degrades performance because the GNN \new{can capture the interaction patterns and fault propagation among instances in microservice systems}, but traditional classifiers cannot understand the graph structure information.

\vspace{-5pt}

\draft{\subsection{Efficiency (RQ3)}}
We record the running time of all methods and compare them in \tabref{table:efficiency}. 
\new{The offline training time of \method is acceptable, particularly when considering its infrequent need for retraining.}
It shows that \method can diagnose one failure within 12 seconds on average online, which means it can achieve quasi-real-time diagnosis because the interval of data collection in \Done and \Dtwo is at least 30 seconds.
Although \method \new{may not possess apparent advantages} among the methods in \tabref{table:efficiency}, \method can meet the needs of online diagnosis.

\vspace{-5pt}

\draft{\subsection{Hyperparameter Sensitivity (RQ4)}}
\label{content:hyperparameter}
\begin{figure*}[!t]
	\centering

\begin{minipage}{\textwidth}
\centering
\includegraphics[]{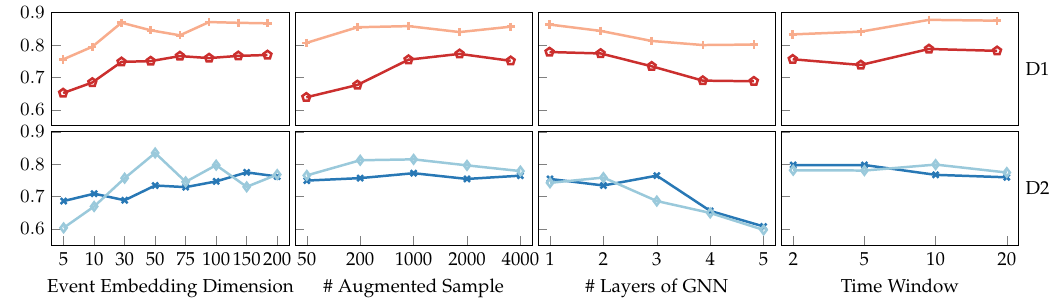}
\end{minipage}

\begin{minipage}{\textwidth}
\centering
    \includegraphics[]{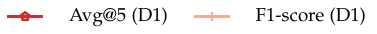}
    \includegraphics[]{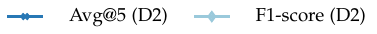}
\end{minipage}

    \vspace{-0.3cm}
    \caption{
        The effectiveness of \method under different hyperparameters.
    }
    \label{fig:parameter-sensitivity}
    \vspace{-0.6cm}
\end{figure*}

We discuss the effect of four hyperparameters of \method.
\figref{fig:parameter-sensitivity} shows how Avg@5 (Task \#1), F1-score (Task \#2) change with different hyperparameters.

\textbf{Embedding dimension.}
The performance of \method reacts differently on different datasets in terms of sensitivity to dimensionality (\Done remains stable while \Dtwo fluctuates more), and the optimal dimensionality is inconsistent across datasets and tasks. We choose the 100 dimensions in our experiments because it has the best overall performance.

\textbf{The number of augmented samples.}
The experiments in \secref{content:eval-effect} show that data augmentation has some improvement in the model's performance. 
However, when the number of samples increases to a certain amount, the information in the training set has already been fully utilized. Instead, the performance may be degraded due to the excessive introduction of noise.
Generally speaking, \method does not need an excessive number of augmented samples as long as the samples are balanced.

\textbf{The number of layers in GNN.}
As the layer number of GNN varies from 1 to 5, the performance of \method in three tasks shows a decreasing trend. 
The model performs best when the layer number is lower than 3.
We do not recommend setting the layer number too large since training deep GNN requires extra training samples, which is hard to meet in real-world microservice systems.

\textbf{Time window.}
The length of the time window has little impact on performance because the moments when failures occur are sparse, and the anomaly events reported in a time window are only relevant to the current failure. 
With accurate anomaly detection, the performance of \method is stable.

\section{Discussion}
\label{sec:discussion}

\subsection{Why Learning-Based Methods?}

The \method approach incorporates several learning-based techniques, such as fastText in the Unified Event Representation (\secref{sec:event-representation}) and GNN (\secref{sec:GNN}).
By doing so, \method significantly outperforms baseline approaches.
We chose to build \method using learning-based methods for the following reasons:
(1) \textit{Accuracy}: learning-based methods provide high accuracy (\secref{sec:evaluation}) and are therefore ideal for diagnosing failures.
(2) \textit{Generalization ability}: failure cases used to train \method contain different patterns of failure propagation for different systems.
A strong generalization ability allows \method to perform robust diagnosis for each system.
(3) \textit{Ability to handle complicated data}: as microservice systems become increasingly complex and monitoring data more high-dimensional, manually setting up rules for failure diagnosis becomes time-consuming and error-prone.
Learning-based methods, on the other hand, take this data as input and learn their relationships, making them well-suited to handle complicated data.

\textbf{Why fastText?}
FastText was chosen because trace, log, and metric data have very different formats.
However, they all share timestamps, meaning they can be sequenced according to their temporal order.
FastText provides superior performance over other static embeddings like word2vec and GloVe, which was demonstrated in \secref{subsec:rq2}.
Although deep dynamic embeddings like ELMo, BERT, and GPT are popular in Natural Language Processing, they are not suitable for microservice settings as the number of failure cases is insufficient to train these large models.

\textbf{Why GNN?}
GNN was chosen because the structure of microservice systems involves many instances and their relationships, which form the structure of a graph.
Various approaches incorporating Random Walk~\cite{ma2020automap, ma2020ms-rank} exist to accomplish failure diagnosis on such graph structures.
However, their ability to generalize is limited since domain knowledge can vary greatly between different systems.
The domain knowledge contained in graph data can be effectively learned by GNNs~\cite{zhang2020deep}, giving them a stronger generalization ability than approaches based on Random Walk.

\textbf{Concerns about learning-based methods.}
While learning-based methods offer several advantages, they do require labeled samples for training.
This can be addressed by (1) utilizing the well-established failure management system in microservice systems as a natural source of failure labeling, (2) \method not requiring too many training samples to achieve good performance (the sizes of \new{the} training set of D1 and D2 are 160 and 80, respectively), and (3) the increasing adoption of chaos engineering, which enables operators to quickly obtain sufficient failure cases.
Several successful practices with the help of chaos engineering have been reported~\cite{zhou2019mepfl, DBLP:journals/tse/ZhouPXSJLD21, yu2021microrank, yang2021aid}.

\subsection{Robustness}
In practice, some modalities can be absent, hindering a successful failure diagnosis system to some extent.
The cause of missing modalities can be generally classified
into three categories.
The first category refers to missing modalities caused by data collection problems.
Modern microservice systems are developing rapidly; the same truth applies to their monitoring agents.
Therefore, it is hard to guarantee that all monitoring data are ideally collected and transmitted.
As a result, missing data is inevitable, which can give rise to missing modalities when specific modalities of the monitoring data are having collection problems.
The second category refers to missing modalities caused by data availability problems.
In some large corporations, monitoring data is individually collected by many different divisions.
Sometimes, specific modalities can be exclusively governed by a division that does not want to disclose its service maintenance data.
Thus, these modalities are collected but not available to general operators.
The third category stands for missing modalities caused by data retrieval problems.
In practice, we often encounter situations where it is very inconvenient to retrieve monitoring data from the data pool.
Multimodal failure diagnosis requires much more data to be collected than single-modal-based methods and may face missing modality problems.
However, an excellent multimodal-based approach should perform well even when some modalities are missing.
We discover that 62 failure cases of \Done lack metric data.
\method is compared with PDiagnose in these cases.
As PDiagnose cannot address Task~\#2, we only present the results of Task~\#1.

\begin{table}[h]
\vspacetbup
\caption{
    \ddraft{Robustness compared to PDiagnose (Task \#1)}
}
\vspacetbup
\label{table:robustness}
\centering
\begin{tabular}{c|cccc}
\toprule
 \multirow{2}[2]{*}{Modality}                 & \multicolumn{2}{c}{\method} & \multicolumn{2}{c}{PDiagnose} \\
                  \cmidrule(lr){2-3} \cmidrule(l){4-5}
                  & A@1         & A@3        & A@1          & A@3          \\ \midrule
Trace, Log, Metric & 0.419        & 0.813       & 0.272         & 0.554         \\
Trace, Log         & 0.274        & 0.661       & 0             & 0.161         \\
\bottomrule
\end{tabular}
\vspacetbdown
\end{table}

As shown in \tabref{table:robustness}, the performance of PDiagnose drops dramatically in these cases, while \method presents salient robustness.
Although \method also witnesses a performance degradation, it is still better than PDiagnose and other Task \#1 baselines.
\method has seen complete data modalities during training and learned a unified representation, allowing it to capture anomalous patterns' correlation to failures better than single-modal-based methods.
On the other hand, PDiagnose treats each modality independently, making it ineffective when facing missing modalities.
\new{To sum up, \method demonstrates robustness since it achieves satisfactory performance even when working with data with incomplete modalities.}

\begin{table*}[!t]
\centering
\caption{
    \new{
    Comparison of \method and existing representative approaches
    }
    }
\label{table:related_work}
    \begin{tabular}{llll}
\toprule
Modality   & Representative Approach                             & Core Technique                        & Diagnosis Result    \\
\midrule
Metric     & AutoMAP\cite{ma2020automap}        & Causal inference \& random walk       & Root Cause Instance \\
Metric     & MS-Rank\cite{ma2020ms-rank}                                            & Causal inference \& random walk       & Root Cause Instance \\
Log        & LogCluster\cite{lin2016logcluster} & Word2vec \& traditional classifier    & Failure Type        \\
Log        & Cloud19\cite{yuan2019cloud19}                                             & Clustering                            & Failure Type        \\
Trace      & MicroRank\cite{yu2021microrank}                                           & Spectrum analysis \& PageRank         & Root Cause Instance \\
Trace      & MicroHECL\cite{jin2021microhecl}   & Graph traverse \& Pearson correlation & Root Cause Instance \\
Multimodal & CloudRCA\cite{zhang2021cloudrca}                                          & Bayesian inference                    & Failure Type        \\
Multimodal & PDiagnose\cite{hou2021pdiagnose}                                           & Vote-based strategy                   & Root Cause Instance \\
Multimodal & \method (ours)                                  & Event embedding \& GNN                   & Root Cause Instance \& Failure Type \\
\bottomrule
\end{tabular}

\end{table*}

\subsection{Concerns about Deployment and Validity}
There are some concerns about deploying \method to real-world microservice systems:
    (1)
    \method needs to adapt to the highly dynamic nature of microservice architecture.
    The stored model of \method can still be effective when service instances are created or destroyed, for \method utilizes the concept of service group as a middle layer.
    The only situation in \new{which} \method needs to be retrained is when new service groups are created.
    However, the creation of service groups is very rare in practice.
    (2)
    Some production systems do not monitor all three modalities at the same time.
    The workflow of \method is general because the event embedding model is trained on event sequences and does not rely on any specific modality.
    Besides, the GNN module deals with feature vectors rather than original monitor data.
    \method can work given that any two of the three modalities are available.

There are two main threats to the validity of the study. 
The first one lies in the limited sizes of the two datasets used in the study. \Done and \Dtwo are relatively smaller than complex industrial microservice systems. The second one lies in the limitation of the failure cases used in the study. Some failure cases of \Done are simpler than industrial failures and represent only a limited part of different types of failures.
However, according to our experiments, \method is effective and robust.
It is very promising that \method can also be effectively applied to much larger industrial microservice systems and more complex failure cases.

\section{Related Work}
\label{sec:related-work}
\textbf{Metric-based failure diagnosis methods.}
Monitoring metrics are one of the most important observable data in microservice systems. 
Many works try to build a dependency graph to depict the interaction between system components during failure, such as Microscope \cite{lin2018microscope}, MS-Rank \cite{ma2020ms-rank}, and AutoMAP \cite{ma2020automap}.
However, the correctness of the above works heavily depends on the parameter settings, which degrades their applicability.
Besides, many methods extract features from system failures, such as Graph-RCA \cite{brandon2020graph-rca} and iSQUAD \cite{ma2020isquad}.
Nonetheless, failure cases are few in microservice systems because operators try to run the system as robustly as possible, severely affecting the performance of these feature-based methods.

\textbf{Trace-based failure diagnosis methods.}
Trace can be used to localize the culprit service, for example,  TraceRCA \cite{li2021tracerca}, MEPFL \cite{zhou2019mepfl}, MicroHECL\cite{jin2021microhecl}, and MicroRank \cite{yu2021microrank}.
\draft{
However, these trace-based methods often focus on the global feature of the systems and do not deal with the local features of a service instance.
}

\textbf{Log-based failure diagnosis methods.}
LogCluster \cite{lin2016logcluster} performs hierarchical clustering on log sequences and matches online log sequences to the most similar cluster.
Cloud19 \cite{yuan2019cloud19} applies word2vec to construct the vectorized representation of a log item and trains classifiers to identify the failure type.
Onion \cite{zhang2021onion} performs contrast analysis on agglomerated log cliques to find incident-indicating logs.
DeepLog \cite{du2017deeplog} and LogFlash \cite{jia2021logflash} integrate anomaly detection and failure diagnosis.
They calculate the deviation from normal status and suggest the root cause accordingly.
\draft{
Log-based methods often ignore the topological feature of microservice systems.
}

\textbf{Multimodal data-based failure diagnosis methods.}
Recently, combining multimodal data to conduct failure diagnosis has drawn increasing attention.
CloudRCA \cite{zhang2021cloudrca} uses both metric and log. It uses the PC algorithm to learn the causal relationship between anomaly patterns of metrics, anomaly patterns of logs, and types of failure. Then it constructs a hierarchical Bayesian Network to infer the failure type.
PDiagnose \cite{hou2021pdiagnose} combines metric, log, and trace. It uses lightweight anomaly detection of the three modalities to detect anomaly patterns. Then its vote-based strategy selects the most severe component as the root cause.
However, these two methods ignore the topology feature of microservice systems.
Groot \cite{wang2021groot} integrates metrics, status logs, and developer activity. 
It needs numerous predefined rules to conduct accurate failure diagnosis, which degrades its applicability to most scenarios.

\new{We compare \method and existing representative approaches in \tabref{table:related_work}.}
In conclusion, compared to single-modal-based methods, \method takes the three important modalities into account. 
Compared to existing multimodal-based methods, \method is among the first to represent different modalities in a unified manner, thus performing more robustly and accurately.

\section{Conclusion}
\label{sec:conclusion}
Failure diagnosis is of great importance for microservice systems. 
In this paper, we first conduct an empirical study to illustrate the importance of using multimodal data (\ie trace, metric, log) for failure diagnosis of microservice systems.
Then we propose \method, an automatic failure diagnosis method, which first extracts events from three modalities of data and applies fastText embedding to unify the event from different modalities.
During training, \method leverages data augmentation to tackle the challenge of data imbalance.
Then it constructs a dependency graph by combining trace and deployment data.
Moreover, \method integrates event embedding and the dependency graph through GNN.
Finally, the GNN reports the root cause instance and the failure type of online failure.
We evaluate \method using two real-world datasets.
The evaluation results confirm the effectiveness and efficiency of \method.

\bibliographystyle{IEEEtran}
\bibliography{main.bib}

\begin{IEEEbiography}[{\includegraphics[width=1in,height=1.25in,clip,keepaspectratio]{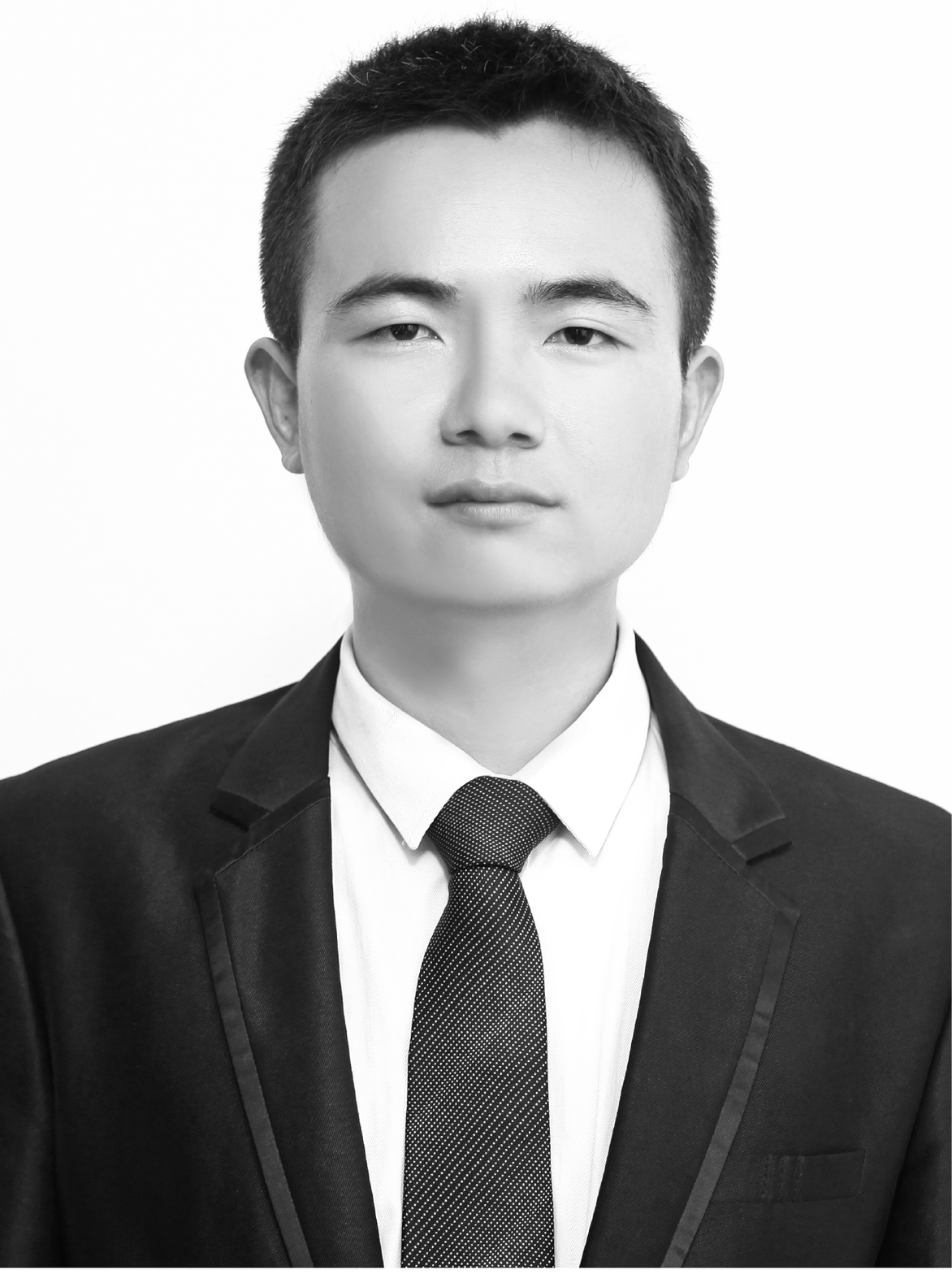}}]{Shenglin Zhang}
    received B.S. in network engineering from the School of Computer Science and Technology, Xidian University, Xi'an, China, in 2012 and Ph.D. in computer science from Tsinghua University, Beijing, China, in 2017. 
    He is currently an associate professor with the College of Software, Nankai University, Tianjin, China.
    His current research interests include failure detection, diagnosis and prediction for service management.
    He is an IEEE Member.
\end{IEEEbiography} 

\begin{IEEEbiography}[{\includegraphics[width=1in,height=1.25in,clip,keepaspectratio]{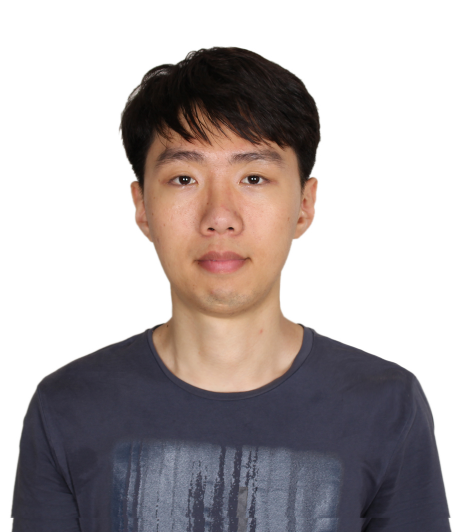}}]{Pengxiang Jin}
    received his bachelor degree in software engineering from Nankai University, Tianjin, China, in 2020. 
    He is currently a master student in the College of Software at Nankai University. His research interests include anomaly detection and anomaly localization.
\end{IEEEbiography} %

\begin{IEEEbiography}[{\includegraphics[width=1in,height=1.25in,clip,keepaspectratio]{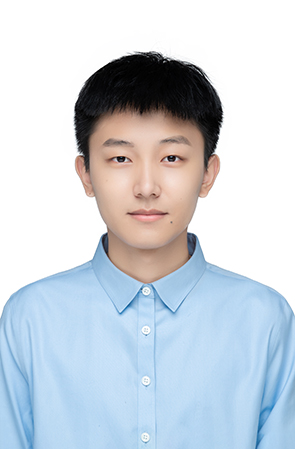}}]{Zihan Lin}
    received his bachelor degree in software engineering from Nankai University, Tianjin, China, in 2021. 
    He is currently a master student in the College of Software at Nankai University. His research interests include failure localization and anomaly detection.
\end{IEEEbiography} %

\begin{IEEEbiography}[{\includegraphics[width=1in,height=1.25in,clip,keepaspectratio]{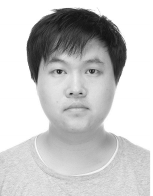}}]{Yongqian Sun}
    received the B.S. degree in statistical specialty from Northwestern Polytechnical
    University, Xi'an, China, in 2012, and Ph.D. in computer science from Tsinghua University, Beijing, China, in 2018. 
    He is currently an assistant professor with the College of Software, Nankai University, Tianjin, China. 
    His research interests include anomaly detection and
    root cause localization in service management.
\end{IEEEbiography} %

\begin{IEEEbiography}[{\includegraphics[width=1in,height=1.25in,clip,keepaspectratio]{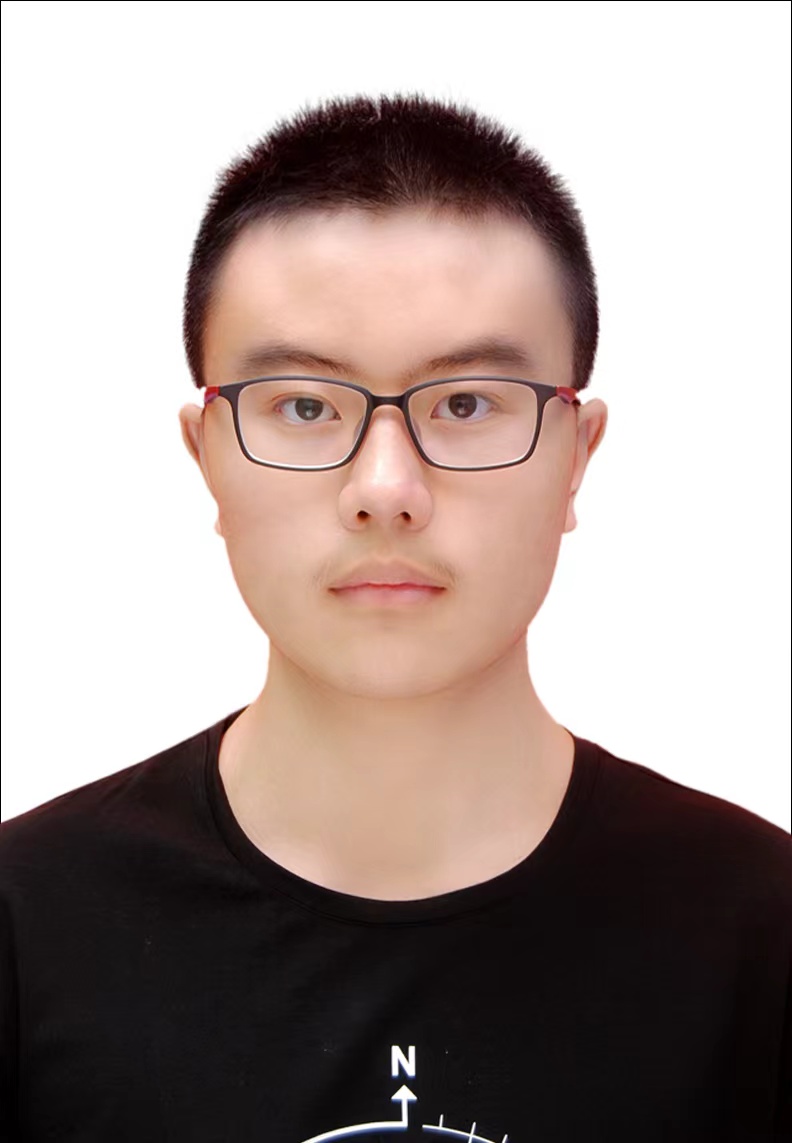}}]{Bicheng Zhang}
    is currently a first-year master student at Fudan University. He received his bachelor degree from Nankai University. His research interests include cloud native and AIOps.
\end{IEEEbiography} %

\begin{IEEEbiography}[{\includegraphics[width=1in,height=1.25in,clip,keepaspectratio]{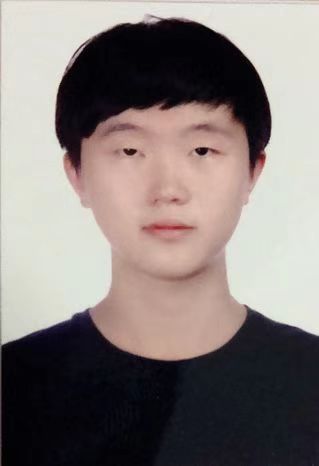}}]{Sibo Xia}
    , born in 2000. Master candidate. 
    His main research interests include knowledge graph, failure detection and diagnosis.
\end{IEEEbiography} %

\begin{IEEEbiography}[{\includegraphics[width=1in,height=1.25in,clip,keepaspectratio]{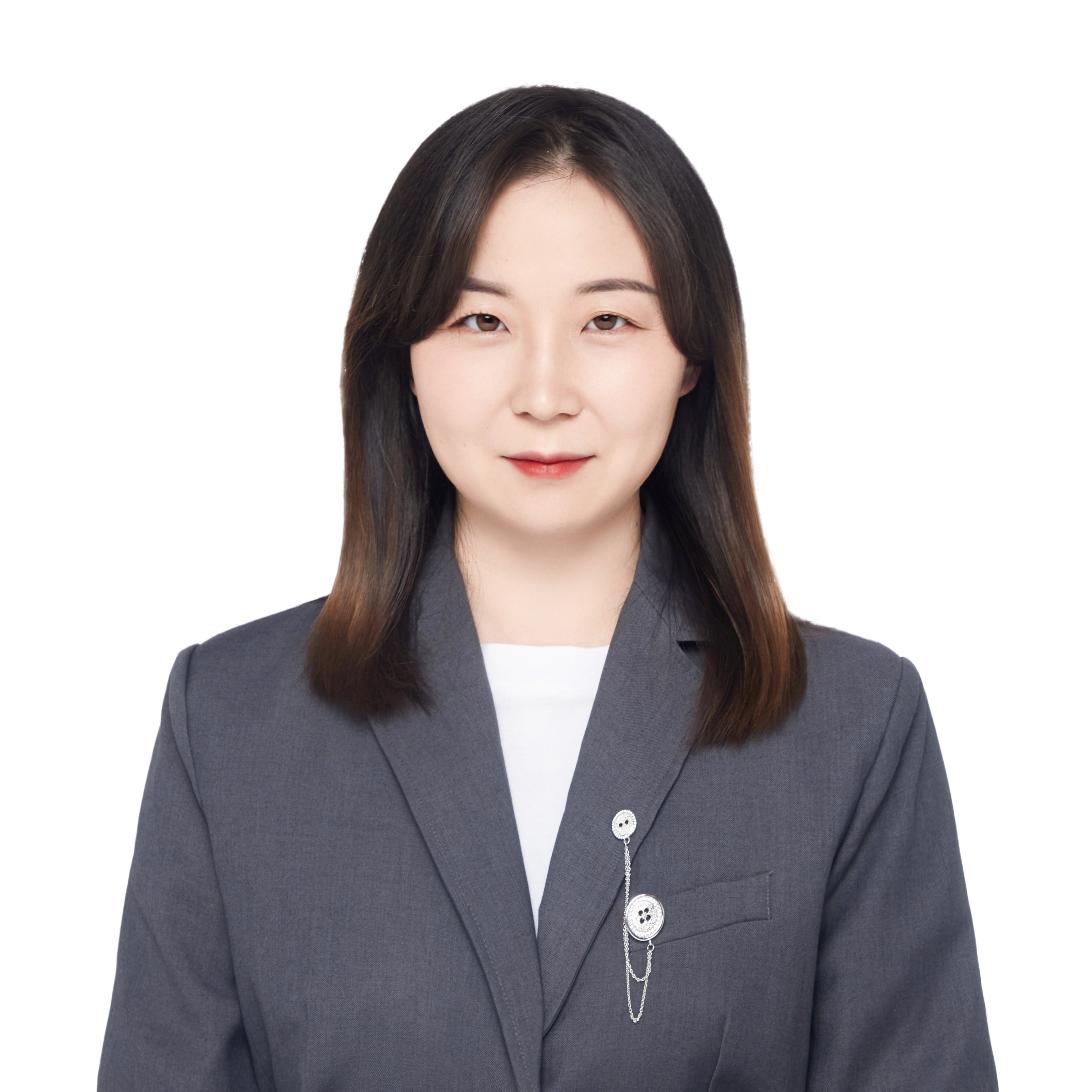}}]{Zhengdan Li}
     is an experimentalist at Nankai University, Tianjin, China. Her research interests include artificial intelligence and software engineering.
\end{IEEEbiography} %

\begin{IEEEbiography}[{\includegraphics[width=1in,height=1.25in,clip,keepaspectratio]{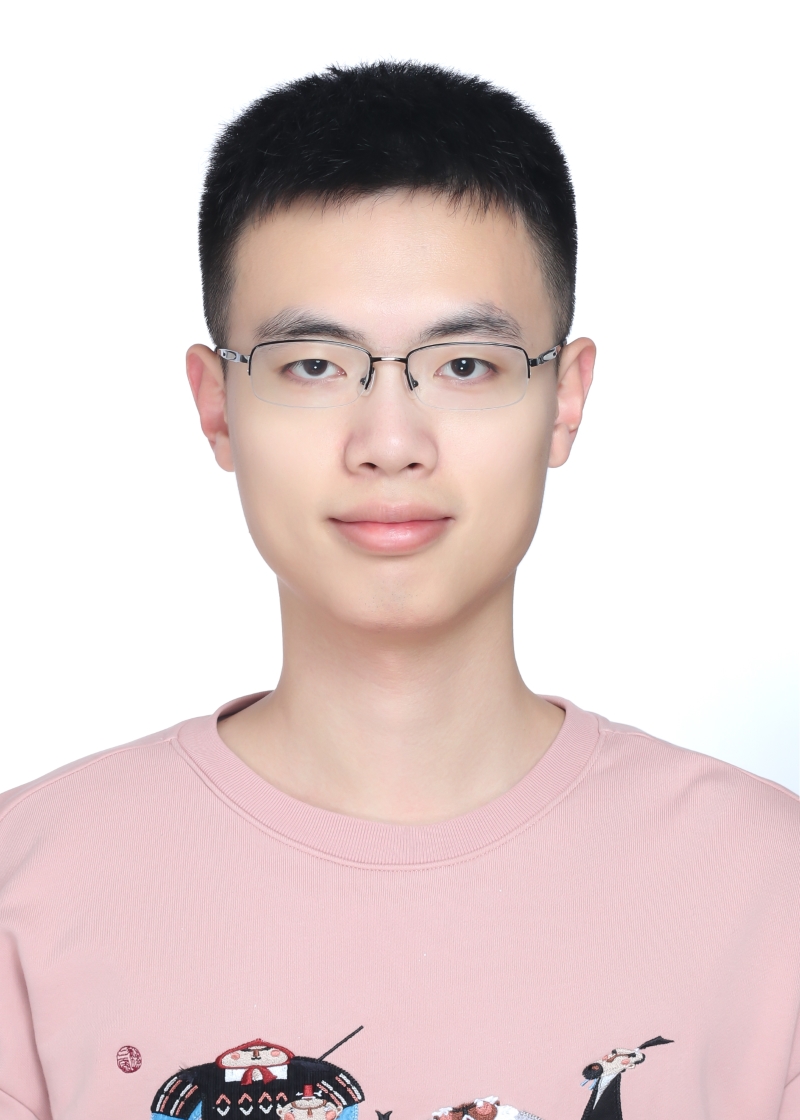}}]{Zhenyu Zhong}
    received his B.S. degree in software engineering from Nankai University, Tianjin, China, in 2020.
    Ph.D. candidate in the College of Software at Nankai University, Tianjin, China.
    His current research interests include anomaly detection, deep learning and NLP.
\end{IEEEbiography} %

\begin{IEEEbiography}
    [{\includegraphics[width=1in,height=1.25in,clip,keepaspectratio]{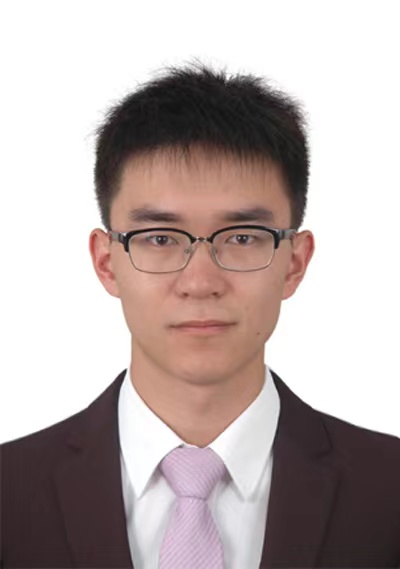}}]{Minghua Ma} 
    is a researcher in Microsoft. Before that, he received his Ph.D. degree from Tsinghua University in 2021. 
    His current research interests include Cloud Intelligence/AIOps.  
\end{IEEEbiography} %

\begin{IEEEbiography}[{\includegraphics[width=1in,height=1.25in,clip,keepaspectratio]{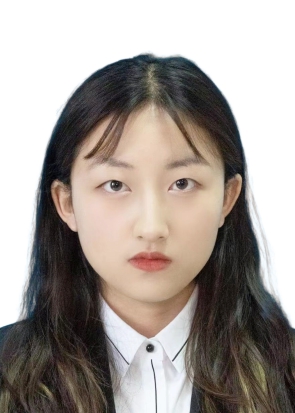}}]{Wa Jin}
    , born in 2001, bachelor candidate. 
    Her main research interests include anomaly detection and failure diagnosis.
\end{IEEEbiography} %

\begin{IEEEbiography}[{\includegraphics[width=1in,height=1.25in,clip,keepaspectratio]{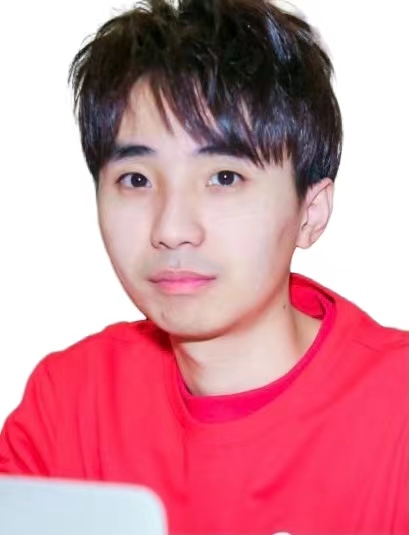}}]{Dai Zhang}
    is employed at ZhejiangE-CommerceBank Co., Ltd., Launched by Ant Group. As a Technical Expert, he mainly focuses on financial basic technical architecture and cloud-native system stability.
\end{IEEEbiography} %

\begin{IEEEbiography}[{\includegraphics[width=1in,height=1.25in,clip,keepaspectratio]{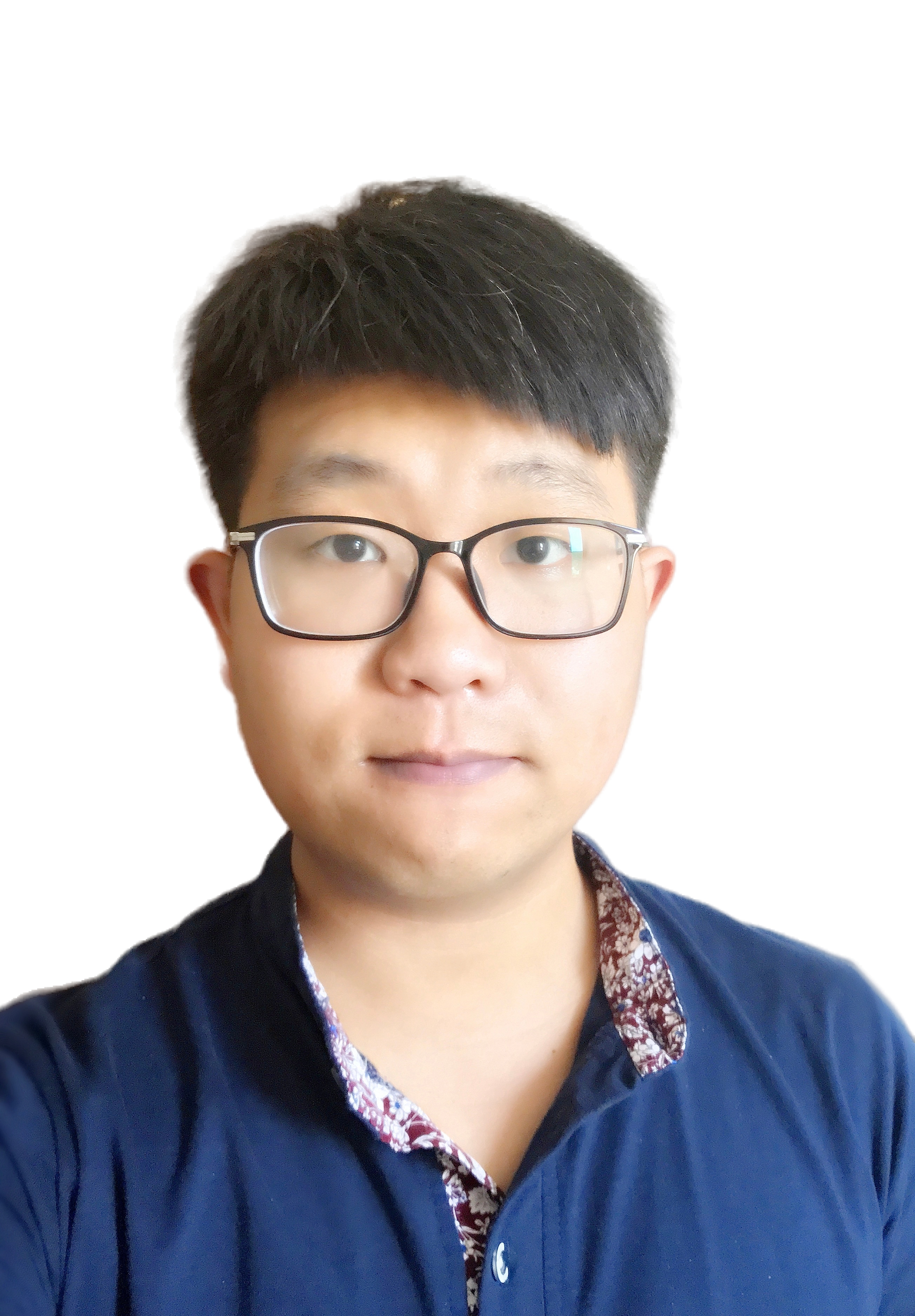}}]{Zhenyu Zhu}
    is employed at ZhejiangE-CommerceBank Co., Ltd., Launched by Ant Group. As a Technical Expert, he mainly focuses on financial basic technical architecture and cloud-native system stability.
\end{IEEEbiography} %

\begin{IEEEbiography}[{\includegraphics[width=1in,height=1.25in,clip,keepaspectratio]{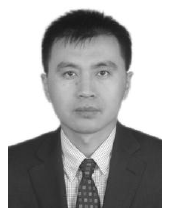}}]{Dan Pei}
    received the B.E. and M.S. degree in computer science from the Department of Computer Science and Technology, Tsinghua University in 1997 and 2000, respectively, and the Ph.D. degree in computer science from the Computer Science Department, University of California, Los Angeles (UCLA) in 2005.
    He is currently an associate professor in the Department of Computer Science and Technology, Tsinghua University.
    His research interests include network and service management in general.
    He is an IEEE senior member and an ACM senior member.
\end{IEEEbiography} %

\end{document}